# Magnitude and Phase-based Feature Fusion Using Co-attention Mechanism for Speaker recognition

Rongfeng Su[1,3], Mengjie Du[4], Xiaokang Liu[1,2], Lan Wang[1,3,B], and Nan Yan[1,3,B]

[1]CAS Key Laboratory of Human-Machine Intelligence-Synergy Systems, Shenzhen Institute of Advanced Technology, Chinese Academy of Sciences, China
[2]University of Chinese Academy of Sciences, Beijing, China
[3]Guangdong-Hong Kong-Macao Joint Laboratory of Human-Machine Intelligence-Synergy Systems, Shenzhen, China
[4]China Telecom Corporation Ltd. Data & AI Technology Company, Beijing, China

**Abstract**. Phase-based features related to vocal source characteristics can be incorporated into magnitude-based speaker recognition systems to improve the system performance. However, traditional feature-level fusion methods typically ignore the unique contributions of speaker semantics in the magnitude and phase domains. To address this issue, this paper proposed a feature-level fusion framework using the co-attention mechanism for speaker recognition. The framework consists of two separate sub-networks for the magnitude and phase domains respectively. Then, the intermediate high-level outputs of both domains are fused by the co-attention mechanism before a pooling layer. A correlation matrix from the co-attention module is supposed to re-assign the weights for dynamically scaling contributions in the magnitude and phase domains according to different pronunciations. Experiments on VoxCeleb showed that the proposed feature-level fusion strategy using the co-attention mechanism gave the Top-1 accuracy of 97.20%, outperforming the state-of-the-art system with 0.82% absolutely, and obtained EER reduction of 0.45% compared to single feature system using FBank.

## 1 Introduction

Speaker recognition is the identification of a person from characteristics of voices. To achieve this goal, most current speaker recognition systems extract a single speaker embedding to represent the speaker's identity from magnitudebased feature inputs, such as FBank and MFCC. According to source-filter theory, speech is the excitation result of vocal tract by vocal source. Thus,

BCorrespondence to: Lan Wang, Shenzhen Institute of Advanced Technology, Chinese Academy of Sciences, China, E-mail: lan.wang@siat.ac.cn; Nan Yan, Shenzhen Institute of Advanced Technology, Chinese Academy of Sciences, China, E-mail: nan.yan@siat.ac.cn.

the latent speaker identity information in speech should be closely related to both magnitude-based features mainly carrying the tract characteristics [1], and phase-based features mainly incorporating the source characteristics [2].

The challenge of integrating both magnitude- and phase-based feature inputs to derive speaker embeddings is the design of appropriate fusion strategies. Existing fusion strategies can be divided into two classes: decision- and feature-level. Decision-level fusion is a simple but effective strategy [3,4] with manually-set weights to separately trained classifiers, which reduces the total risk of accepting an incorrect hypothesis. Nonetheless, the separate training procedure ignores the interrelationship between the magnitude and phase domains. To capture the deep interrelationship, feature-level fusion strategies can be used. Specifically, researchers directly concatenate the output of a hidden layer of the magnitudebased and phase-based models through joint training [5–8]. A key issue associated with existing feature-level fusion strategies is the equal contribution assumption of arbitrary pronunciation in the magnitude and phase domains. These methods assign the same weights to the magnitude- and phase-based features for different speech contents in the fusion stage. However, the reliability of speaker features represented by the amplitude and phase domains should be distinct for different speech contents. Therefore, for a given speech, the representations with more speaker discriminative power should have higher weights.

To address this issue, this paper proposed a novel feature-level fusion framework using the co-attention mechanism for speaker recognition. The framework consists of two separate sub-networks for the magnitude and phase domains, respectively. The intermediate high-level outputs of both domains are fused by the co-attention mechanism before a pooling layer. Inspired by the successful application of co-attention in Video Question Answering [9,10], the co-attention module generates a correlation matrix to capture the speaker semantics from both domains and reassigns weights for dynamically scaling contributions based on different pronunciations in both domains. The proposed co-attention-based fusion strategy gave the Top-1 accuracy of 97.20% on the VoxCeleb1 SID subtask [11] outperforming the state-of-the-art system [12] with 0.82% absolutely, and obtained EER reduction of 0.45% compared to single feature system using FBank for SV subtask.

## 2   Phase-based Feature Extraction

Phase-based features reflect vocal source characteristics, such as pitch and harmonic peaks [1,4]. However, the origin phase spectrum is difficult to be used due to the phase wrapping phenomenon. To obtain robust phase representations, various phase-related feature extraction methods have been developed, such as group delay [13–16], residual phase and instantaneous frequency [16]. The *modified group delay* (MODGD) [13] is used as the phase-related feature.

The original group delay $\tau(\omega)$ is defined as the negative derivative of phase spectrum without phase unwrapping,

$$\tau(\omega) = -\frac{d\theta(\omega)}{d\omega} = -\text{Im}(\frac{d\log X(\omega)}{d\omega}) \quad (1)$$

where $X(\omega)$ is the corresponding short-time Fourier transformation (STFT) of a given speech signal sequence $\{x[n]\}$, $\theta(\omega)$ is the phase spectrum of $X(\omega)$ and Im( ) means the imaginary part of $\log X(\omega)/d\omega$ in Equation (1). It equals,

$$\tau(\omega) = \frac{X_R(\omega)Y_R(\omega) + X_I(\omega)Y_I(\omega)}{|X(\omega)|^2} \quad (2)$$

where $Y(\omega)$ is the STFT of the sequence $\{nx[n]\}$, the subscripts $R$ and $I$ denote the real and imaginary parts of the complex spectrum, respectively.

The original group delay is meaningful and can resemble the magnitude spectrum, only if the speech signal is a minimum phase signal [13]. Otherwise, it becomes spiky and unstable around formants [17]. From Equation (2), $\tau(\omega)$ will have a sharp increase in value, when the denominator $|X(\omega)|^2$ tends close to zero. It arises from the proximity of zero points to the unit circle in *Z-transform* sight. MODGD $\tau_m(\omega)$ smoothens undesired spikes by suppressing zeros into the unit circle radially [13],

$$\tau_m(\omega) = \tau(\omega)|\tau(\omega)|^{\alpha-1} \quad (3) \quad \text{s.t.}$$

$$\tau(\omega) = \frac{X_R(\omega)Y_R(\omega) + X_I(\omega)Y_I(\omega)}{|S(\omega)|^{2\gamma}} \quad (4)$$

where $\alpha = 0.4, \gamma = 0.9$ reported in [18] and $S(\omega)$ is the cepstral smoothed version of $X(\omega)$. Unlike [13,17,18], DCT-II is omitted. We use standardized $\tau_m$ as MODGD directly for clear harmonic preservation and local correlation modeling.

## 3 Magnitude- and Phase-based Baseline Systems

### 3.1 Network Architecture

For the speaker embedding extraction, the TDNN-based x-vectors [19–22] and the ResNet-based models [23–26] have achieved dominant performance in recent years. In this work, we use Thin ResNet34 [24,25] to extract speaker embeddings from either FBank or MODGD inputs, since it retains the feature extraction capability of the original ResNet34 [23] at the same depth with lower computational cost. The self-attentive pooling [24] is applied to aggregate framelevel speaker embeddings to the utterance-level as well. The details of the Thin ResNet34 are shown in Table 1.

### 3.2 Loss Functions

*Cross Entropy Loss*:

$$L_1 = -\log \frac{e^{\mathbf{W}_{y_i}^T \mathbf{x}_i + b_{y_i}}}{\sum_{j=1}^{C} e^{\mathbf{W}_j^T \mathbf{x}_i + b_j}} \quad (5)$$

Table 1. The Architecture of Thin ResNet34. Each row of the table specifies the convolutional kernel size, channel numbers and stride step. *SAP* denotes the self-attentive pooling layer.

| layer name | Thin ResNet34 | |
|---|---|---|
| | MODGD | Fbank |
| Conv0 | $7 \times 7, 16, 2 \times 1$ <br> $7 \times 1, 16, 3 \times 1$ | $7 \times 7, 16, 2 \times 1$ |
| ResBlock1 | $\begin{pmatrix} 3 \times 3, 16 \\ 3 \times 3, 16 \end{pmatrix} * 3, 1 \times 1$ | |
| ResBlock2 | $\begin{pmatrix} 3 \times 3, 32 \\ 3 \times 3, 32 \end{pmatrix} * 4, 2 \times 2$ | |
| ResBlock3 | $\begin{pmatrix} 3 \times 3, 64 \\ 3 \times 3, 64 \end{pmatrix} * 6, 2 \times 2$ | |
| ResBlock4 | $\begin{pmatrix} 3 \times 3, 128 \\ 3 \times 3, 128 \end{pmatrix} * 3, 1 \times 1$ | |
| SAP | - | |

where $x_i$ is vector of *i*-th speaker, $y_i$ is the $y_i$-th class, C is the class number, W and *b* are the learnable weights of the last classification layer.

*AAM-Softmax* loss [27] introduces additive angular margin penalty:

$$L_2 = -log \frac{e^{s\,cos(\theta_{y_i}+m)}}{e^{s\,cos(\theta_{y_i}+m)} + \sum_{j=1, j \neq y_i}^{C} e^{s\,cos\theta_j}} \qquad (6)$$

where $cos(\theta_y)$ is the cosine similarity of x and W, and *s* and *m* are the scale factor and penalty margin, respectively.

## 4  Fusion Strategies

### 4.1  Decision-level Fusion

In the decision-level fusion, the final similarity score is calculated with a manually-set ratio $0 \leq r \leq 1$,

$$s_i^d = r \times s_i^g + (1-r) \times s_i^f \qquad (7)$$

where $s^g_i$, $s^f_i$, $s^d_i$ denote the similarity score of the *i*-th test speaker from the MODGD system, the FBank system and the combined decision-level fusion result, respectively. The ratio *r* is set to 0.5 in this paper. Supposed that *N* is the number of the enrolled speakers, the cosine similarity score $s_i \in R^N$ of the *i*-th test speaker is:

$$s^i = \frac{1}{M} \sum_{j=1}^{M} \text{cosine}(\boldsymbol{e}_{i,j}, \boldsymbol{E}) \qquad (8)$$

where *M* is the number of the utterances of the *i*-th test speaker, $e_{i,j} \in R^D$ denotes the utterance-level speaker embedding of the *j*-th utterance of the *i*-th speaker, and $E \in R^{N \times D}$ denotes the enrolled speaker embedding database.

### 4.2  Feature-level Fusion

The feature-level fusion framework consists of two parts: feature extractors and the feature fusion module. Since magnitude- and phase-based features characterize different physical properties of a speaker, two parallel feature extractors without shared parameters are used to obtain the uniqueness of speaker semantics latent in the magnitude and phase domains.

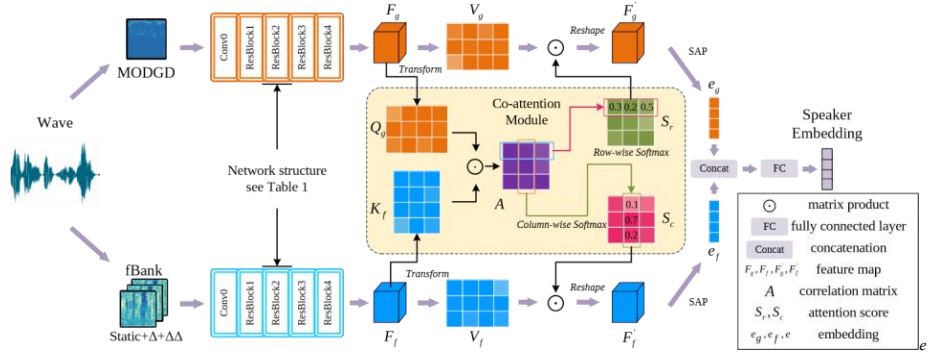

Fig.1. The proposed feature-level fusion framework with the co-attention mechanism. For a given speech input, the correlation matrix $A$ re-assigns the weights and thus dynamically regularizes the contributions in the magnitude and phase domains.

Traditional Concatenation When the co-attention module in Fig.1 is removed, the rest part is the traditional feature-level fusion. It is an effective fusion strategy after feature projection. Through joint training, it captures the underlying interaction between the magnitude and phase domains in the common feature space. To reduce the computational cost, we apply concatenation at the embedding level. The final speaker embedding $e \in \mathbb{R}^D$ is:

$$e = W_\psi^T [W_{\phi 1}^T e_f, W_{\phi 2}^T e_g] \tag{9}$$

where $e_g, e_f \in \mathbb{R}^D$ are self-attentive pooled MODGD and FBank speaker embeddings, respectively, $W_\psi^T \in \mathbb{R}^{2D \times D}$, $W_{\phi 1}^T, W_{\phi 2}^T \in \mathbb{R}^{D \times D}$ are the transformation function weights, and $[\cdot]$ is the concatenation operation.

Co-attention Mechanism Compared to the phase-based features, the magnitude based features should have different ability to characterize speaker identity for various pronunciation. Thus, for a given speech, the proposed channel coattention mechanism is supposed to encode the correlation between them, and to constrain the contribution of each domain to the speaker embedding. In this case, the co-attention mechanism functions as a cross-domain dynamic adjustment of weights. The details of the framework is illustrated in Fig 1. The core correlation matrix $A \in \mathbb{R}^{C \times C}$ is calculated as:

$$A = Q_g K_f^T = F_g W_1 (F_f W_2)^T \tag{10}$$

where $F_g, F_f \in R^{C \times H \times W}$ are the feature maps from the adjacent feature extractors, $Q_g, K_f \in R^{C \times HW}$ are the flattened transformation of feature maps, $W_1, W_2$ are the learnable convolution weights, and the subscripts $f$ and $g$ denote FBank and MODGD, respectively. $A_{ij}$ ($1 \leq i,j \leq C$) is the similarity score between the $i$-channel of feature map MODGD and $j$-th channel of FBank. Attention score $S_c$ and $S_r \in R^{C \times C}$ reflecting the relevance are normalized with column-wise and row-wise softmax function,

$$S_i = \begin{cases} \text{softmax}(A^T) &, i = c \\ \text{softmax}(A) &, i = r \end{cases} \tag{11}$$

The adjusted and re-emphasized feature maps $F_f'$ and $F_g' \in \mathbb{R}^{C \times H \times W}$ are obtained,

$$F_f' = \text{Reshape}(S_c V_f) \qquad F_g' = \text{Reshape}(S_r V_g) \tag{12}$$

where $V_f, V_g$ are the convolution transformation of original feature maps, and *Reshape* is a function that transforms a tensor of $R^{C \times HW}$ to the shape of $R^{C \times H \times W}$.

## 5 Experiments

### 5.1 Dataset

VoxCeleb [11] is currently the most popular open-source dataset in the field of speaker recognition. For SID, we conducted experiments on the official split of VoxCeleb1, where train and dev contain 145,265 utterances of 1,251 speakers for training, and test contains 8,251 utterances of the same speakers for test. For speaker verification, the entire VoxCeleb2 was used as training set and the official split in VoxCeleb1 was used for test.

### 5.2 Network Inputs

In our experiments, 64-dimensional static FBank features, with the corresponding delta and double delta coefficients, were extracted by Kaldi toolkit with the 25ms window and 10ms shift. Utterance-level cepstral mean variance normalization was applied as well. 201-dimensional MODGD features were obtained through the algorithm described in Section 2 with the 25ms window and 10ms shift. In addition, a

3s-sliding window with 1s shift was applied for each utterance as time augmentation in training stage.

### 5.3 Implementation Details and Evaluation Metrics

Our experiments were based on Pytorch framework. We trained all systems on 4 NVIDIA RTX A6000 GPUs with a fixed batch size of 64. Adam was used as the optimizer during training stage, with an initial learning rate of 1e-4 and a weight decay factor of 0.05 for each epoch. Besides the *Softmax* loss function, *AAM-Softmax* was also considered to be used in this paper. For *AAM-Softmax*, we set *margin* = 0.2 and *scale* = 30. We used *Top-1 Accuracy* as the SID evaluation metric, *Equal Error Rate* and *minDCF* for SV. The two *FAR, FRR* weights in *minDCF* were set as 1, and the priori probability of a target speaker was set as 0.05.

Table 2. Speaker identification subtask: the performance of various speaker recognition systems trained on VoxCeleb1. *S* in *Loss* column denotes *Softmax*, *A-S* denotes *ASoftmax*, and *AAM-S* denotes *AAM-Softmax*, respectively.

| System | Input | Fusion strategy | | Loss | Acc(%) |
|---|---|---|---|---|---|
| ¬VGG-M [11] | Spectrogram | - | | S | 80.50 |
| - ResNet [24] | FBank | - | | A-S | 89.90 |
| ® ResNet [28] | Spectrogram | - | | S | 89.00 |
| ¯Transformer[12] | FBank | - | | - | 96.38 |
| °-I | FBank | - | | S | 91.61 |
| °-II | | - | | AAM-S | 96.48 |
| °-III | MODGD | - | | S | 89.13 |
| °-IV | | - | | AAM-S | 94.88 |
| °-V | FBank +MODGD | decision-level | | S | 93.13 |
| °-VI | | | | AAM-S | 96.89 |
| °-VII | | feature -level | Traditional | S | 94.00 |
| °-VIII | | | | AAM-S | 97.04 |
| °-IX | | | Co-attention | S | 94.96 |
| °-X | | | | AAM-S | 97.20 |

## 6 Results and Analysis

### 6.1 Speaker Identification Subtask

Table 2 shows performance comparison between different systems.Five trends can be generalized from these results.

Magnitude-based vs Phase-based. Phase-based features can also be used to characterize speaker identity for speaker recognition. For example, the Top-1 accuracy of system °-III using MODGD was only 2.71% lower than the FBank baseline °-I.

With vs Without Fusion. All fusion Systems (line 5 to 14) consistently outperformed the comparable baseline systems °-III, °-I using single feature inputs.

For example, in Table 2, system °-V using decision-level fusion with *Softmax*, even inferior to other fusion strategies, gave a higher Top-1 accuracy of 1.52% absolutely over the FBank baseline °-I. *Acc* of either system °-VII or °-IX in Table 2 are higher than those of systems with only FBank or MODGD.

Decision-level vs Feature-level Fusion. Feature-level fusion strategies might be more suitable than decision-level fusion strategies to integrate both features. Traditional vs Co-attention Feature-level Fusion. Compared with the traditional feature-level fusion strategy, the proposed co-attention mechanism can further improve the system performance by re-assigning weights automatically for different representations of speech content. As shown in Table 2, system °-IX with co-attention mechanism outperformed system °-VII using the traditional feature-level fusion strategy by Top-1 accuracy of 0.96% absolutely. The same system °-IX with co-attention in Table 2 achieved more accuracy of 0.38% than °-VII on unseen speaker test.

Softmax vs AAM-Softmax. Training with more competitive *AAM-Softmax* can dramatically improve SID system. For example, the proposed system °-X from Fig.1 with *AAM-Softmax* gave the best performance of 97.2%, and outperformed the state-of-the-art result [12] by Top-1 accuracy increasing of 0.88% absolutely.

Table 3. Speaker verification subtask: the performance of various speaker recognition systems in Table 2 trained on VoxCeleb2.

| System | Input | Feature-level Fusion Strategy | Params(M) | Loss | EER(%) | minDCF |
|---|---|---|---|---|---|---|
| °-I | FBank | - | 1.39 | S | 4.26 | 0.289 |
| °-II | | | | AAM-S | 2.37 | 0.152 |
| °-III | MODGD | - | 1.42 | S | 4.40 | 0.295 |
| °-IV | | | | AAM-S | 2.45 | 0.163 |
| °-V | FBank+ MODGD | Traditional Co-attention | 2.97 | S | 4.18 | 0.268 |
| °-VI | | | | AAM-S | 2.26 | 0.138 |
| °-VII | | | 2.98 | S | 3.81 | 0.252 |
| °-VIII | | | | AAM-S | 2.04 | 0.132 |

## 6.2 Speaker Verification Subtask

As shown in Table 3, the similar trends in SID subtask could also be found on SV subtask. For example, compared with the speaker recognition systems using single feature inputs (from °-I to °-IV in Table 3), the speaker recognition systems using feature-level fusion strategy (from °-V to °-VIII in Table 3) gave lower EER and minDCF. In addition, the speaker recognition system using the proposed co-attention mechanism outperformed the corresponding speaker recognition system using the traditional fusion method. For example, when using *AAM-Softmax* as the loss function, an EER reduction of 0.22% absolute (9.7% relative) was obtained from the system °-VIII over the system °-VI.

## 7 Conclusions

In this paper, we proposed a novel feature-level fusion framework using the co-attention mechanism for speaker recognition. Experiments demonstrated that by modeling the correlation between magnitude and phase domains, the contributions from the two domains to the speaker semantics are automatically scaled according to different speech contents, which makes an impressive improvement for speaker recognition systems. Since MODGD still requires additional data processing, this makes the whole system complex. In our future work, we will focus on end-to-end methods for extracting magnitude- and phase-based features.

## 8  Acknowledgement

This work is supported by National Key R&D Program of China (U23B2018), Shenzhen Science and Technology Program (JCYJ20220818101411025, JCYJ202 20818101217037, JCYJ20210324115810030), Shenzhen Peacock Team Project (KQTD20200820113106007), and National Natural Science Foundation of China (NSFC 62271477).